\documentclass[conference]{IEEEtran}
\let\labelindent\relax
\usepackage{cite}
\usepackage{amsmath,amssymb,amsfonts}
\usepackage{algorithmic}
\usepackage{algorithm}
\usepackage{graphicx}
\usepackage{textcomp}
\usepackage{xcolor}
\usepackage[hyphens]{url}
\usepackage{fancyhdr}
\usepackage{multirow}
\usepackage{multicol}
\usepackage{enumitem}

\ifCLASSINFOpdf
\else
\fi
\begin{document}
%
\title{Dishonest Approximate Computing: A Coming Crisis for Cloud Clients}

\author{\IEEEauthorblockN{Ye Wang, Jian Dong, Ming Han and Jin Wu}
\IEEEauthorblockA{Harbin Institute of Technology\\
\{yewang, dan, minghan, wujin\}@hit.edu.cn}
\and
\IEEEauthorblockN{Gang Qu}
\IEEEauthorblockA{University of Maryland\\
gangqu@umd.edu}
}

\maketitle

\begin{abstract}
Approximate Computing (AC) has emerged as a promising technique for achieving energy-efficient architectures and is expected to become an effective technique for reducing the electricity cost for cloud service providers (CSP). However, the potential misuse of AC has not received adequate attention, which is a coming crisis behind the blueprint of AC. Driven by the pursuit of illegal financial profits, untrusted CSPs may deploy low-cost AC devices and deceive clients by presenting AC services as promised accurate computing products, while falsely claiming AC outputs as accurate results. This misuse of AC will cause both financial loss and computing degradation to cloud clients. In this paper, we define this malicious attack as DisHonest Approximate Computing (DHAC) and analyze the technical challenges faced by clients in detecting such attacks. To address this issue, we propose two golden model free detection methods: Residual Class Check (RCC) and Forward-Backward Check (FBC). RCC provides clients a low-cost approach to infer the residual class to which a legitimate accurate output should belong. By comparing the residual class of the returned result, clients can determine whether a computing service contains any AC elements. FBC detects potential DHAC by computing an invertible check branch using the intermediate values of the program. It compares the values before entering and after returning from the check branch to identify any discrepancies. Both RCC and FBC can be executed concurrently with real computing tasks, enabling real-time DHAC detection with current inputs. Our experimental results show that both RCC and FBC can detect over 96\%-99\% of DHAC cases without misjudging any legitimate accurate results.
\end{abstract}


%

\section{Introduction}

Approximate computing (AC) is an energy-efficient design methodology which takes advantage of the error-resistant feature in modern applications such as neural networks, and improves energy efficiency by sacrificing part of computing accuracy but on the premise of acceptable output qualities. Its significant optimization on energy efficiency makes AC a potential technique to break the energy- and power-wall of the Moore’s Law. Some research outcomes have already made their impact on industry, such as Google's tensor processing unit (TPU) in deep learning chips and the S1 chip developed by Singular Computing and DARPA \cite{liu2018approximate}. For cloud service providers (CSP), AC devices has been an attractive option for reducing electricity costs while offering additional selectable pricing schemes \cite{armeniakos2022hardware}. Furthermore, in addition to traditional accurate computing services at the standard charge, CSPs can provide low-cost approximate computing services at a discounted price, catering to the clients who have limited budgets and do not mind AC errors \cite{wang2020approximate}. This flexibility allows clients to choose the desired service level based on their specific computing quality requirements. For example, clients engaged in scientific computing or neural network training can choose discounted AC service, obtaining approximate results which may not be perfect but are still acceptable in terms of quality.

\begin{figure*}[htb]
    \centering
    \includegraphics[scale=0.55]{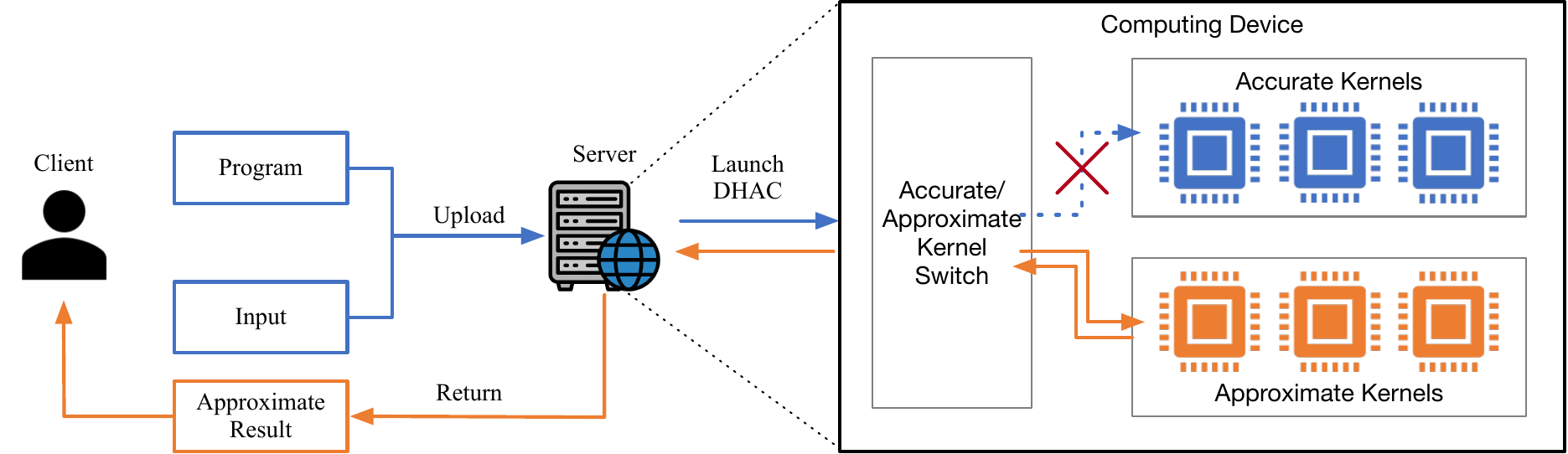}
    \caption{Dishonest Approximate Computing}
    \label{fig:Untrusted Approximate Computing}
\end{figure*}

However, behind the promising blueprint of AC, there exists a concerning possibility of malicious misuse. An untrusted CSP can deceive clients by using AC devices or kernels to serve the clients who pay for accurate service at the standard fee, while falsely claiming that they have provided the promised accurate computing products. The primary motivation behind this attack is to gain illegal financial benefits. This malicious activity can be seen as a form of commercial fraud as the CSP misrepresents lower-cost and lower-quality products (AC services) as standard products (accurate computing services). In this paper, we define this malicious attack launched by CSPs as DisHonest Approximate Computing (DHAC), as illustrated in Fig. \ref{fig:Untrusted Approximate Computing}. DHAC not only leads to direct financial losses for the clients but also degrades the quality of the received results. Firstly, clients pay the standard fee for accurate services but only receive lower-priced AC services in return, which leads to a financial loss for the clients. Secondly, the AC errors will degrade the overall quality of computation. Although the AC error may not be significant enough to completely crash the client’s application, the client should have the rightful expectation to receive accurate results with a higher level of quality. Moreover, AC errors can negatively affect the convergence rate of iteration-based applications such as neural network training. Consequently, the server may take longer to complete the computing task, leading to increased costs for the clients as they have to extend the rental period.

Detecting DHAC poses several challenges and disadvantages for clients. Apart from limitations in clients' local computing resources, clients face a lack of visibility into the inner architectures of the server's devices. This allows CSPs to easily manipulate the system configuration inventories and conceal the true underlying hardware from users. Additionally, the diverse range of AC methods also makes it difficult to get any prior knowledge about the specific AC technique the server may utilize. To detect DHAC, validating the results through repetitively executing the clients’ computation tasks shall not be a practical approach because it significantly increases the users' costs. While simple test code snippets or small test programs may be effective when the server solely uses AC architectures, they may not be sufficient in more complex scenarios where the server employs more sophisticated strategies. In this paper, we aim to address DHAC in such complex scenarios where straightforward tests may not effectively combat such attacks. Specifically, we consider scenarios where CSPs can selectively switch tasks between accurate and AC devices. The malicious server strategically launches DHAC for significant computing tasks, while accurately executing small programs that the server deems as potential honest tests. Furthermore, during the initial deployment of significant programs, the server disguises itself as a trustworthy entity for a period of time before initiating DHAC. This cautious approach is taken because users are more likely to scrutinize the results during the early stages of deployment. These actions are not difficult to implement for the server. From the client’s perspective, the server’s sophisticated strategy of launching DHAC creates more challenges. Merely attempting to detect the server alone with test programs cannot guarantee that the real computation tasks will be free from DHAC at a later stage. Whether the client uses a small test program or executes the entire computing task during the initial deployment and compare results, the server can bypass these detections and launch DHAC after the client stops detecting and starts the real computing tasks. In summary, clients require a low-cost detection method which can be integrated with any program, perform DHAC detection while the real computing task is being executed, and determine whether the current result is computed accurately or approximately with any given input.

To effectively address the potential threat of DHAC, the design of the detection method should consider the specific requirements and constraints of clients. First, the detection method should be cost-effective and have low computing resource requirements. Clients typically have limited resources for executing the entire computing task. Otherwise, they will not rent a server for computing purposes. Secondly, the detection method should not rely on any golden model as a reference. Clients cannot directly infer the expected result with real-world inputs, making it unrealistic to assume the availability of a golden model.Instead, the clients need a method which can detect DHAC while justifying whether the result is accurate or approximate given the current input. Thirdly, in addition to achieving high DHAC detection accuracy, the method should maintain a low false-positive rate. Mistakenly identifying a significant number of legitimate accurate results as DHAC will negatively impact the execution efficiency of the real computing tasks. Each time the detection method reports a DHAC, the client needs to recompute for a correct result and further investigate whether or not the CSP is performing malicious behavior.

In this paper, our primary target is to highlight the potential risk associated with the misuse of AC techniques and propose effective detection methods to combat DHAC attacks. We propose two detection methods, namely Residual Class Check (RCC) and Forward-Backward Check (FBC), which aim to assist clients in overcoming the challenges of DHAC detection. Both RCC and FBC are designed to be executed concurrently with real computing tasks, enabling real-time determination of whether the server’s returned result is accurate or approximate. These two methods do not rely on any golden model and do not require any prior knowledge of expected output values or the server's hardware architecture. Additionally, the client can implement low-cost monitoring of the candidate server over an extended period of time. Specifically, RCC offers the advantage of being imperceptible and allows for online adjustments, while FBC provides broader applicability and shifts the check cost to the server rather than burdening the clients. By proposing these detection methods, our aim is to provide clients with effective tools to combat DHAC attacks.

Our main contributions can be summarized as follows:

\begin{enumerate}[leftmargin=*]
    \item In contrast to the focus on developing low-power AC techniques, this paper explores the dark side behind AC’s blueprint. We analyze the high possibility and serious consequences when AC is illegally misused driven by the pursuit of illegal financial profit, and propose a malicious scenario called DHAC, wherein a server deceives its clients by presenting AC services as promised accurate computing products.
    \item We propose RCC and FBC as two DHAC detection methods. RCC allows clients to infer the residual class to which a legitimate accurate output should belong, enabling clients to detect the presence of AC elements by comparing the returned result’s residual class. FBC starts from the intermediate values of the program and detects possible DHAC by computing an invertible check branch. It compares the values before entering and after returning from the check branch to identify any discrepancies. Once a real-time result successfully passes through the RCC or FBC check, it can be confidently considered as an accurate result, providing clients with assurance to use it without concerns. These two methods offer distinct advantages in terms of applicability, obscurity, independency and cost. Both of them can detect DHAC attacks individually and are not mutually exclusive, allowing for their collaborative use.
    \item We simulate a DHAC attack scenario and evaluate the detection accuracy of the two proposed methods, RCC and FBC. According to our experimental results, both RCC can FBC can detect over 96\%-99\% of DHAC results without any misjudgment of legitimate accurate results.
\end{enumerate}

The rest of this paper is organized as follows: In Section II, we summary the backgrounds and related works of AC. In Section III, we mainly propose the threat model of DHAC and state some assumptions about the ables and disables of clients and servers. In Section IV and Section V, we propose the working principles and detection processes of the two DHAC detect methods, RCC and FBC, respectively. The experimental results are presented and analyzed in Section VI and we finally conclude our work in Section VII.

\section{Background and Related Work}

The essence behind AC is trading computation accuracy for lower power consumption. The roots of AC can be traced back to the 1960s \cite{mitchell1962computer}. Nowadays, the significant advancements and widespread application of AI technology further highlight the value of AC. On one hand, many modern AI tasks inherently possess error resilience, making them well-suited to benefit from the introduction of AC. Applications such as machine translation, signal processing, and object detection can tolerate certain levels of AC errors without significant negative impacts on their computation quality. On the other hand, many modern AI applications face challenges related to high energy consumption and large computing resource requirements. Many CSPs can offer various compute-intensive products, but the associated electricity costs are considerably high. Therefore, there is an urgent need for high energy-efficient designs, in which area AC has excellent performance, to enable AI tasks to be effectively utilized on battery-powered devices and reduce energy costs for CSPs \cite{armeniakos2022hardware}.

\subsection{AC Mechanisms}

In the past decade, there has been extensive research exploring the feasibility of approximation in logic circuits and arithmetic units \cite{barua2019approximate}. Adders and multipliers, as the most crucial components in the computing process, have been the primary focus of AC research \cite{mittal2016survey}. Some AC circuit designs, such as error-tolerant adder type II (ETAII) \cite{zhu2010enhanced}\cite{zhu2009design}, generic accuracy configurable adder (GeAr) \cite{shafique2015low}, carry cut-back adder (CCBA) \cite{camus2016low}, ApproxLP multiplier \cite{imani2019approxlp}, partial product perforation multiplier (PPAM) \cite{zervakis2016design}, and inaccurate counter-based multiplier (ICM) \cite{lin2013high}, have shown the significant ability to optimize energy consumption \cite{jiang2020approximate}. The authors in \cite{mrazek2017evoapprox8b} and \cite{mrazek2018scalable} utilized genetic algorithms to systematically explore the design space of approximate arithmetic units and established a fundamental library of approximate circuits called EvoApproxLib. This library includes numerous examples of approximate adders and multipliers, which can assist in designing approximate computing platforms. Due to significant energy consumption optimization, AC is transcending typical design approaches \cite{rapp2021mlcad} and has emerged as a new design paradigm for energy-efficient devices and accelerators. An approximate architecture which combines various approximation designs will lead to a new trend in future computer design. Some research projects continuously develop AC architectures, including Eyeriss chip for deep learning developed by MIT \cite{chen2019eyeriss}, Neurostream chip for high performance neural network \cite{azarkhish2017neurostream}, and BitFusion for approximate GPU \cite{sharma2018bit}. Some industry products, such as the tensor processing unit (TPU) in Google’s deep learning chip and DARPA’s Singular Computing chip \cite{liu2018approximate}, have already incorporated AC techniques can contributed to real-world computing tasks. It is foreseeable that mature AC devices will enter the market in the future, especially benefiting CSPs by significantly reducing electricity costs.

\subsection{Security Issues of AC}

As the development of AC progresses from laboratory design to industrial deployments, some researches have started to explore the potential drawbacks and security concerns associated with AC architectures. Authors in \cite{rahmati2015probable} primarily discussed privacy leakage in AC modules. They highlight how the unique error distribution of AC modules can potentially reveal the identity of users. Authors in \cite{yellu2019security} and \cite{regazzoni2018security} demonstrated how traditional security risks, such as hardware Trojan, side channel attack and reverse engineering, can pose threats to AC architectures. In \cite{yellu2020security} and \cite{yellu2022securing}, the authors illustrated how hardware Trojans endanger AC and proposed countermeasures against such attacks. In \cite{wang2020approximate}, the authors presented how error injection and data modification trigger an uncontrollable error to an AC system. In \cite{ahmed2023maas}, the authors discussed the potential of hardware Trojans insertion during the approximate accelerator synthesis. The authors in \cite{yellu2023inead} proposed a method to distinguish between inaccuracies caused by approximation and those caused by attacks. While existing researches mainly focus on traditional vulnerabilities, this paper introduces DHAC, a new threat caused by misusing which arises with the widespread application of AC techniques.

\subsection{DHAC and SLA Violation}

DHAC and malicious Service Level Agreement (SLA) violation share some similarities as they are both driven by illegal financial gains. However, they differ in their techniques. Malicious SLA violation involves intentionally reducing promised computation resources, such as CPU speed, memory frequency, and storage space, below the agreed-upon service level with clients. CSPs may illegally decrease the average costs for clients while serving more clients through virtualization and resource pooling techniques, thereby generating additional rental income. The authors in \cite{houlihan2014auditing} pointed out that by providing users with fewer resources, CSPs can support more users on the same hardware and increase their profits. To address this malicious scenario, the authors presented a scheme for auditing SLA violations on VM's CPU speed. In \cite{zhang2014verifying} and \cite{ye2012verifying}, the authors proposed a third-party auditor based framework to test the virtual machines on the server, enabling the detection of potential malicious SLA violations on VMs' memory size. On the other hand, DHAC is more complex in terms of technique because attackers can use both hardware and software AC methods to launch such attacks. This means that DHAC can occur at various design levels depending on the specific AC techniques implemented by the CSP. Meanwhile, controlled by CSPs, AC devices can possess similar physical characteristics as traditional accurate devices, such as latency and response speed, which make it hard to detect DHAC through side channels.

\section{Assumptions and Threat Model}

In this section, we state some assumptions regarding the capabilities and limitations of servers and clients. Then we build a threat model of DHAC to describe the intentions of malicious CSPs and the requirements of clients.

\subsection{Assumptions}

\textbf{CSP (the attacker, DHAC launcher):}

\begin{figure}
    \centering
    \includegraphics[scale=0.46]{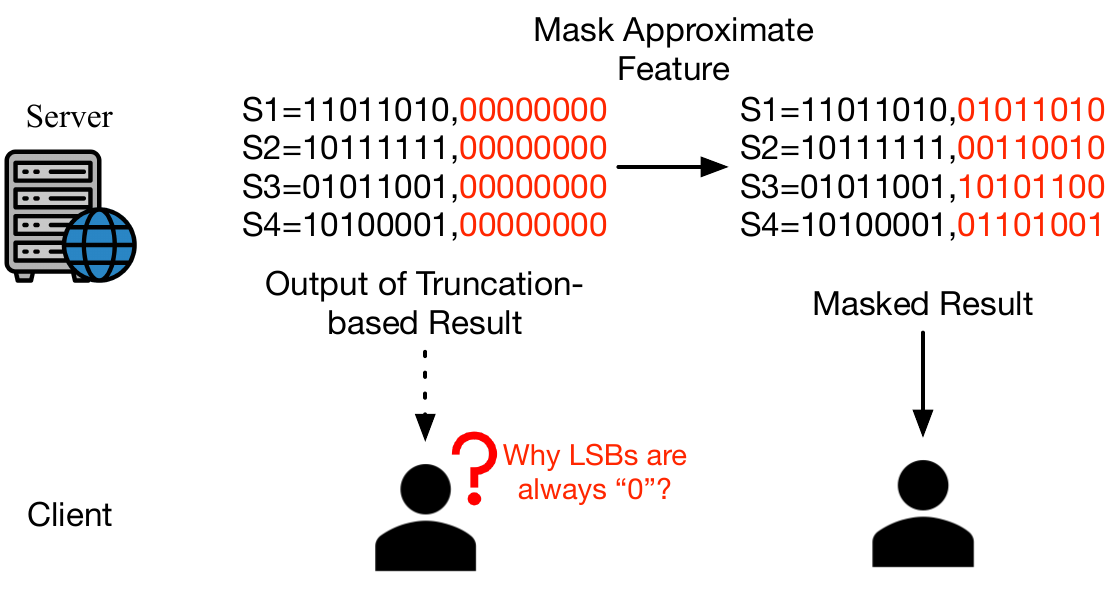}
    \caption{Approximation Feature Masking}
    \label{fig:Approximate Feature Masking}
\end{figure}

\begin{enumerate}[leftmargin=*]
    \item The CSP deploys both traditional accurate computing devices and approximate hardware based AC devices or kernels. It has the flexibility to switch between the accurate computing paradigm $P_{acc}$ and the AC paradigm $P_{appx}$ in different executions. 
    \item The CSP only performs DHAC when it determines that the expected benefits outweigh the potential risk of being detected. Any small programs or code snippets that the server perceives as potential honest tests will be executed accurately. The CSP also has the capability to actively switch back from $P_{appx}$ to $P_{acc}$ when it suspects that clients may be monitoring its behavior or checking the program.
    \item The CSP is unable to thoroughly analyze each program of every user. After launched DHAC, the CSP will perform $P_{appx}$ on the entire program uploaded by the client. 
    \item The approximation error of $P_{appx}$ is always carefully controlled by the server. Even though the CSP have launched DHAC, the client will never receive obvious erroneous results or intuitively sense the presence of DHAC.
    \item The server has the capability to directly modify the format or value of the output to conceal any AC features. The client cannot point out the existing of AC according to the output data's characteristics, as shown in Fig. \ref{fig:Approximate Feature Masking}.
    \item The server can easily manipulate and forge the system configuration information, thereby concealing the utilization of AC components. 
\end{enumerate}

\textbf{Client (the victims):}

\begin{enumerate}[leftmargin=*]
    \item The client has full access to the program before it is sent to the server and has full access to the results returned by the server.
    \item The server’s real hardware architecture remains a black box to the client. The client is unaware of whether the server is executing $P_{acc}$ or $P_{appx}$ in a given execution.
    \item The client does not possess any golden model of their program or any prior information about the accurate result. Specifically, for any given input, the client cannot predict the value, value range, or any patterns of the accurate result without executing the entire program.
\end{enumerate}

\subsection{Threat Model}

In the context of DHAC, the CSP is the attacker. The primary objective of launching DHAC is to seek illegal financial benefits by deceiving clients. The CSP accomplishes this by disguising lower-price and lower-quality AC services as promised accurate computing products. While performing DHAC, the CSP has the ability to switch its computing paradigm between $P_{acc}$ and $P_{appx}$ to evade possible detection from the client's side.

The main victims are the clients who rent accurate services. The main requirement for the client is a low-cost, low false positive rate, and golden model free detection method to help monitor possible DHAC attacks in long-term services or detect DHAC during any specific execution of the uploaded program. The detection method should be able to be integrated with the main computation tasks, and can determine whether a real-time result provided by the server is computed accurately or approximately for any given input.

\section{Residual Class Check}

In this section, we propose the principle and workflow of Residual Class Check (RCC). The design of RCC is motivated by the mathematical principle that the correctness of any arithmetic computing process can be verified by recomputing the process within a residual class ring. By using this principle, RCC enables the checking process to be executed using energy-efficient simple operations within a controllable bit width, making it affordable for clients to verify or monitor the server’s behavior over an extended period of time.

\subsection{Mathematical Foundations}
We first reference the definition and basic arithmetic operations, addition and multiplication, of residual class in algebra as the mathematical foundation of this section:

\textbf{Definition 1: }For two integers $m$ and $r$, where $r>0$ and $0\le r<m$, the integer set
$$
\bar{r}=\{q m+r \mid q \in Z\}
$$
is defined as a \textit{Residual Class} module $m$.

\textbf{Definition 2: }For an integer $m>0$, we denote the set of all the residual classes module $m$ as:
$$
Z_m=\{\overline{0}, \overline{1}, \cdots, \overline{m-1}\}
$$
For $\bar{a}, \bar{b} \in Z_m$, we define the addition and multiplication of residual class module $m$ as:
$$
\bar{a} \oplus \bar{b}=\overline{a+b}, \quad \bar{a} \otimes \bar{b}=\overline{a \times b}
$$
The residue class set $Z_m$, together with the addition $\oplus$ and the multiplication $\otimes$ is defined as a Residual Class Ring module $m$, denoted as $\mathbb{Z}_m$

Obviously, any residual class ring $\mathbb{Z}_m$ has a \textit{zero element} $\bar{0}$ and a \textit{unit element} $\bar{1}$, because for any $\bar{a} \in Z_m$, we have $\bar{a} \oplus \bar{0}=\bar{a}$ and $\bar{a} \otimes \bar{1}=\bar{a}$. We further incorporate the following two theorems to extend the addition and multiplication to all four basic arithmetic operations. In the interest of brevity, we omit the proofs of the following two theorems here, as they belong to the fundamentals of algebra.

\textbf{Theorem 1: }For an integer $m>0$ and any $\bar{a}, \bar{b} \in Z_m$, we have:
$$
\bar{a} \otimes \bar{b}=\bar{b} \otimes \bar{a}
$$

\textbf{Theorem 2: }If $m>0$ and $m$ is a prime number, for any $\bar{a} \in Z_m$, we can find $\bar{b} \in Z_m$ which satisfies:
$$
\bar{a} \otimes \bar{b}=\bar{b} \otimes \bar{a}=\overline{1}
$$
where $\bar{b}$ is referred as the \textit{inverse} of $\bar{a}$ and denoted as $\bar{b} = \bar{a}^{-1}$

Therefore, by selecting a prime number $m$, a client can map any arithmetic computing process $F(\cdot)$ from the linear space to the corresponding computing process $F_m(\cdot)$ within the residual class ring $\mathbb{Z}_m$. The basic arithmetic operations in $\mathbb{Z}_m$ can be transferred as follows:
\begin{equation}
\label{4-1}
    \left\{\begin{array}{c}
    a+b \mapsto \bar{a} \oplus \bar{b} \\
    a-b \mapsto \bar{a} \oplus(-\bar{b}) \\
    a \times b \mapsto \bar{a} \otimes \bar{b} \\
    \frac{a}{b} \mapsto \bar{a} \otimes \bar{b}^{-1}
    \end{array}\right.
\end{equation}
where $a$ and $b$ are both integers and $\bar{a}, \bar{b} \in \mathbb{Z}_m$. With any input $I=$ $\left(i_1, i_2, \cdots i_n\right)$, we have:
$$
F\left(i_1, i_2, \cdots i_n\right) \equiv F_m\left(\overline{l_1}, \overline{l_2}, \cdots \overline{l_n}\right) \quad \bmod m
$$
which is the basic mathematical principle of RCC.

Operations performed within the residual class ring provide significant efficiency advantages compared to those in the linear space. This efficiency is achieved through the utilization of operations with shorter and controllable bit widths, which are determined once the module $\textit{m}$ is defined and are independent of the original bit width in the linear space. By leveraging the principles of residual class rings, we can design a low-cost method to check the server’s behavior without the need for executing the entire program with accurate computations. Clients can infer the correct residual class to which a legitimate result should belong. Any result generated by approximate paradigms will, with a high probability, fall into an incorrect residual class because of the approximation errors. 

\subsection{Check Process}

\begin{figure*}
    \centering
    \includegraphics[scale=0.48]{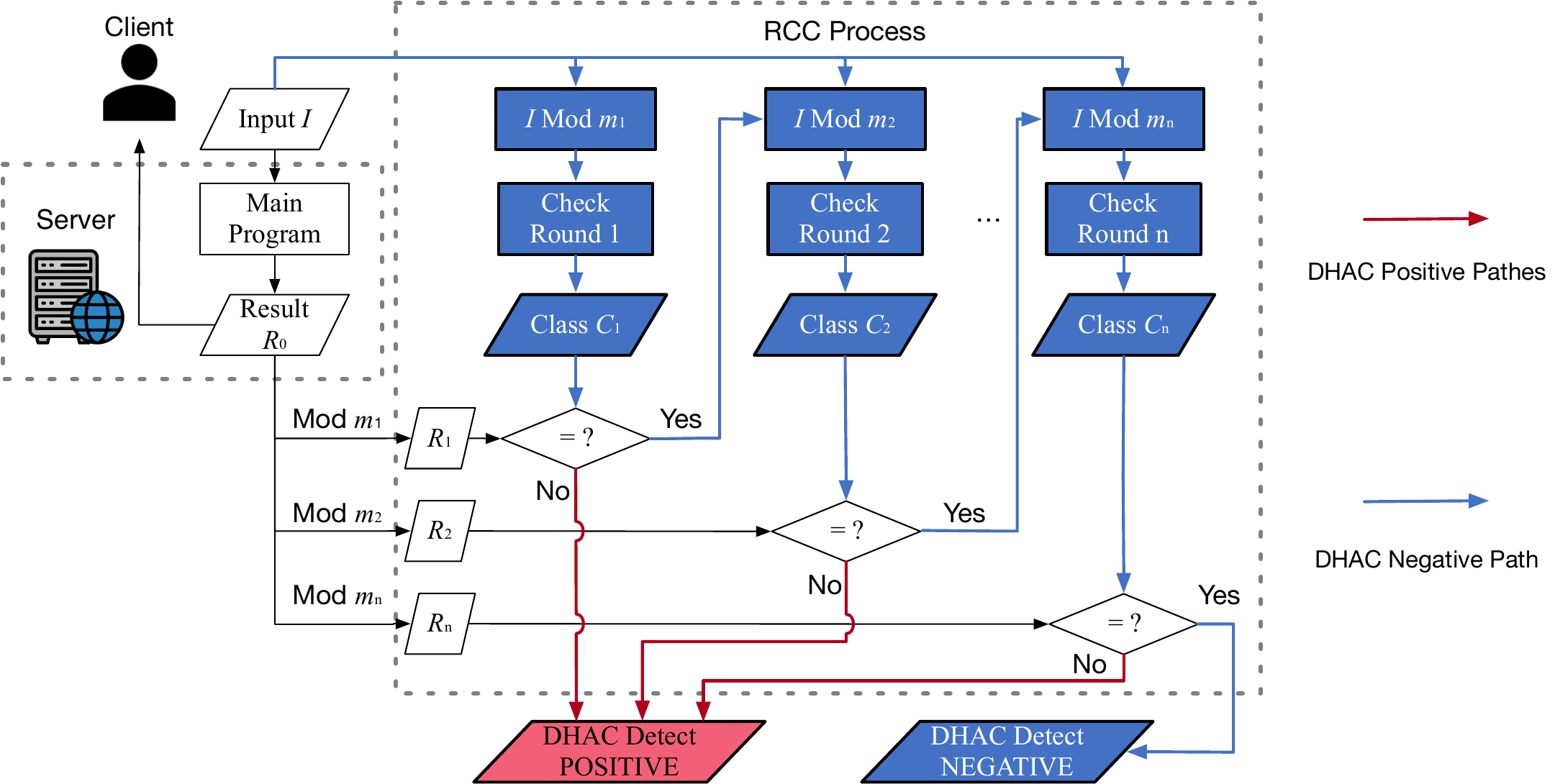}
    \caption{The Workflow of RCC Process}
    \label{fig:The Workflow of RCC Process}
\end{figure*}

RCC is a low-cost, multi-round checking process. This approach is locally executable and independent with the server’s computing process. The workflow of an RCC process is shown in Fig. \ref{fig:The Workflow of RCC Process}.

First, the client uploads the original input $I$ and the main program to the server. The server then performs the computation task and return the result $R_0$ to the client. However, the client is unaware of whether $R_0$ was computed using $P_{a c c}$ or $P_{a p p x}$. To initiate the RCC process, the client needs to settle a set of modules $M=\left\{m_1, m_2, \ldots, m_q\right\}$. According to Theorem 2, if the program does not involve division operations, the module set can be chosen from arbitrary integers. Otherwise, all elements in the module set must be prime numbers. For each $m_i \in M$, the RCC process executes an independent check round, as illustrated in Algorithm \ref{Algorithm 1}. Both the original inputs $I$ and the candidate result $R_0$ from the linear space are transformed into their respective residual class, denoted as $\bar{I}_i$ and $\overline{R_{0 i}}$, with current module $m_i$. The RCC process then follows the same data flow as the main program but start with $\bar{I}_i$, and all arithmetic operations are performed using the corresponding operations in the residual class ring, as defined in formula \ref{4-1}. During the check round for module $m_i$, the RCC process determines the correct residual class $\bar{C}_i$ to which the accurate result should belong. If at any round the check process occurs $\overline{R_{0 i}} \neq \bar{C}_i$, the RCC process indicates that the server may have launched a DHAC attack. Only when all rounds of check have passed, that is, for any $i$, we have $\overline{R_{0 i}}=\bar{C}_i$, can it be concluded that the server is not suspected of launching a DHAC attack, and the current result can be trusted.

\begin{algorithm}
\caption{Multi-Round RCC Process}
\label{Algorithm 1}
\hspace*{0in}{\textbf{Input:}} Input $I=\left\{i_1, i_2, \ldots, i_p\right\}$\\
Check module set $M=\left\{m_1, m_2, \ldots, m_q\right\}$\\
Candidate server's result $R$\\
\hspace*{0in}{\textbf{Output:}} DHAC judgement result $J \in \{$Positive, Negative$\}$

\begin{algorithmic}[1]
\STATE {$J \leftarrow Negative$}
\FOR{$m_j$ in $M$}
    \STATE {$\bar{I} \leftarrow $ [ ] }
    \STATE {$\bar{R} \leftarrow (R ~\%~ m_j)$}
    \FOR{$i_k$ in $I$}
        \STATE {$\bar{I}$.append $(i_k ~\%~ m_j)$}
    \ENDFOR
    \STATE {$C_j \leftarrow F_m(\bar{I})$}
    \IF{$C_j \neq R_j$}
        \STATE {$J \leftarrow$ Positive}
        \STATE {\textbf{Break}}
    \ENDIF 
\ENDFOR
\STATE {return $J$}
\end{algorithmic}

\end{algorithm}

\subsection{Check Segment Extract}

When implementing an RCC process, the client has the flexibility to extract specific code segments rather than using the entire original program for the check rounds. Our evaluations in Section VI demonstrate that a small code segment with tens of arithmetic operations can effectively detect DHAC attacks. For complex applications with millions of operations, selecting a longer code segment or directly using the entire original program would only increase the detection cost without providing additional benefits. Moreover, in certain real-world applications, the output may not have mathematical meanings. For instance, in a CNN used for image classification tasks, the output represents a classification result and does not correspond to a numerical value. In such cases, the client needs to carefully select the arithmetic operation parts of the CNN, such as the convolution kernel, as the code segment to be included in the check rounds. This ensures that the selected segments have mathematical relevance and can be effectively utilized for the DHAC detection process.

RCC offers flexibility in selecting program segments, without imposing specific requirements on their structure. Clients can divide the applications into main steps or stages, such as filters in image processing applications or kernels in neural networks. The client can then choose the computation-intensive steps as the check segments for the RCC process. The only additional setting required is to insert two intermediate value exports at the beginning and the end of the selected segments. This allows for the capture of inputs used in the checking process and the candidate result $R_0$.

\begin{figure}
    \centering
    \includegraphics[scale=0.35]{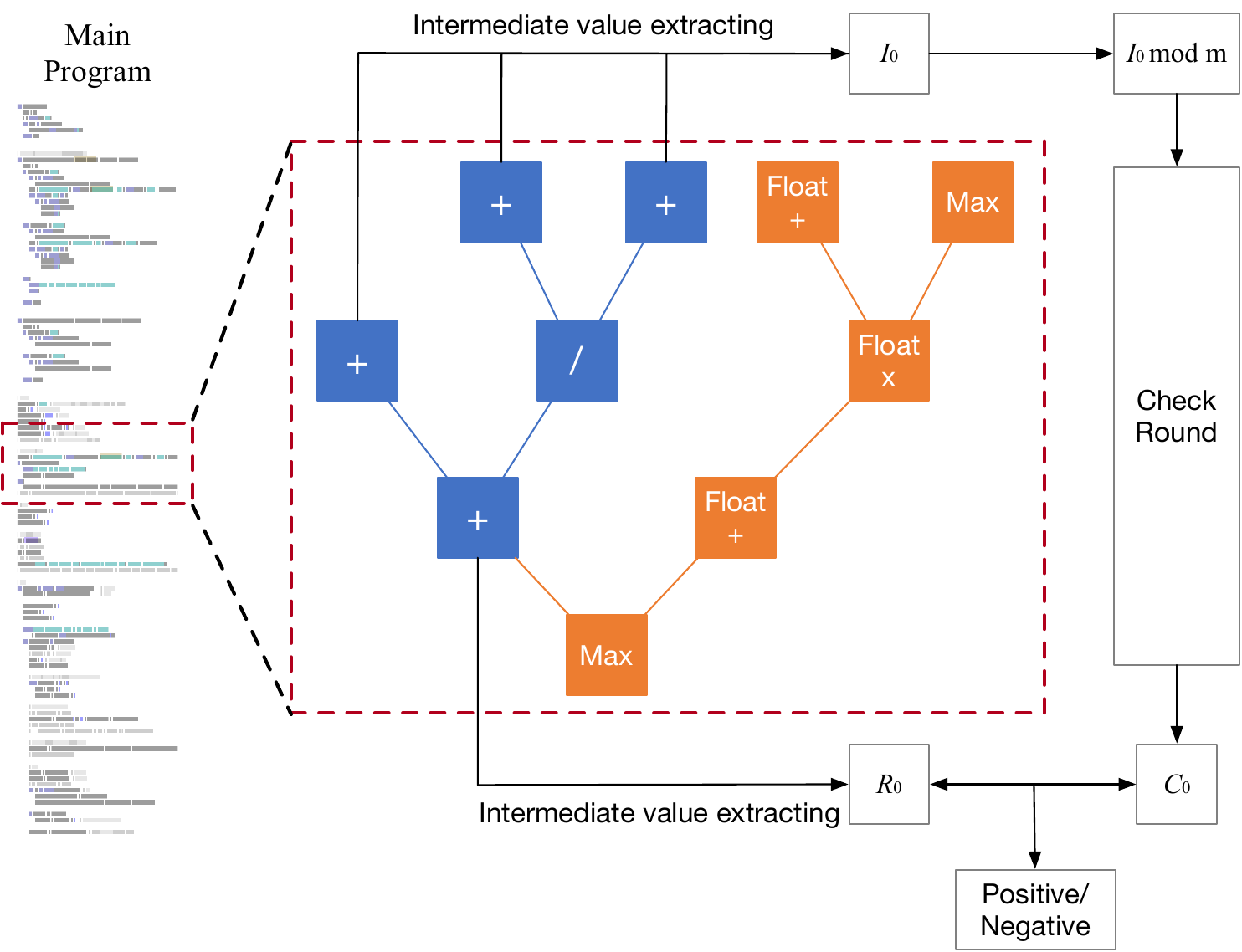}
    \caption{Check Segment Extracting}
    \label{fig:Check Segment Extracting}
\end{figure}

Alternatively, clients can utilize a breadth-first search-based method to select code segments suitable for RCC at the data flow graph level, as shown in Fig. \ref{fig:Check Segment Extracting}. The search algorithm traverses the data flow graph, starting with integer operations as initial points. It progressively expands the subgraphs originating from these initial points until either all child nodes become non-integer arithmetic operations or the client-specified maximum depth is reached. The subgraphs with appropriate depth will be used in the check rounds, with two intermediate value exports inserted at their beginning and ending.

By extracting check segments, clients can apply RCC to any programs which incorporate integer arithmetic components. Although a client can execute the check segment directly in the linear space after extraction, RCC can further reduce the detection cost by utilizing a shorter bit width, or perform more/longer segments with the same computing resources. However, RCC has a limitation in its exclusive applicability to integer operations due to its underlying mathematical principles. It cannot by directly applied to programs where floating-point computations are prevalent. In the next section, we will introduce Forward-Backward Check (FBC), which overcomes the constraint of RCC and enables detecting DHAC in floating-point programs.

\section{Forward-Backward Check}

In DHAC scenarios, one significant drawback faced by the client is the lack of prior knowledge regarding the expected outputs. It is challenging for the client to determine whether an output contains any AC errors based solely on its value. To address this issue, we propose FBC, which is a program instrumentation-based check method. 

The core principle of FBC is to instrument several sentinel code snippets, whose outputs can be predicted with certainty under accurate computing paradigms, into the original computing tasks. Through these sentinel branches, FBC establishes several predictable anchor points into the program. Consequently, if the CSP launches DHAC, the approximate computing error will lead to deviations between the computed results and the expected values, and thereby triggering an alert from the sentinel.

In this section, we will illustrate the workflow of FBC, and further discuss the main advantages and disadvantages compared with RCC.

\subsection{FBC Process}

\begin{figure*}
    \centering
    \includegraphics[scale=0.4]{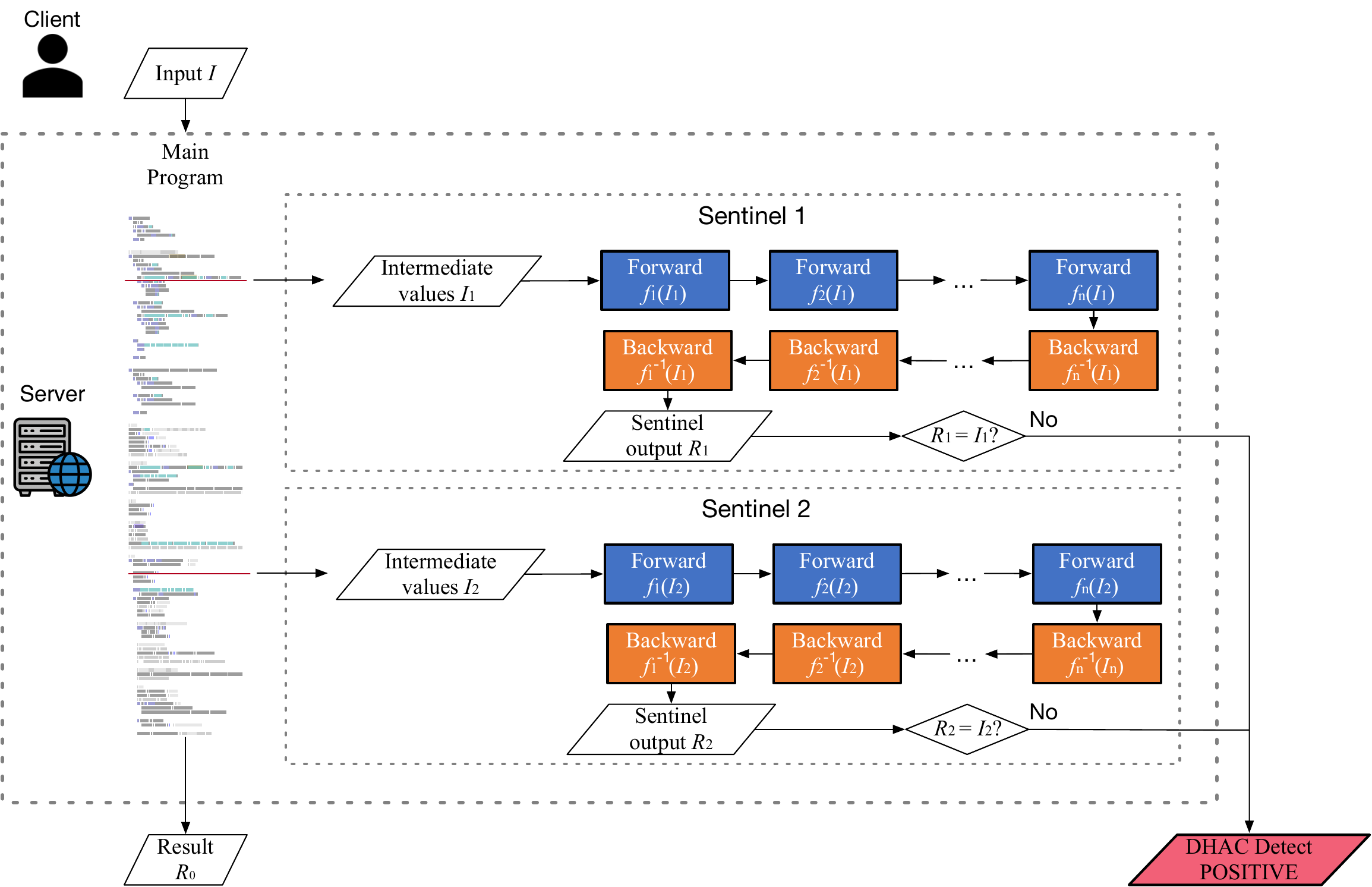}
    \caption{The Workflow of FBC Process}
    \label{fig:The Workflow of FBC Process}
\end{figure*}

Fig. \ref{fig:The Workflow of FBC Process} shows the workflow of FBC process. In this example, the client instruments two sentinels into the original computing task. From the program's perspective, a sentinel is a code branch which does not contribute to the actual computation but serves exclusively for the purpose of DHAC detection. During the instrumentation phase, a data entrance is created at the branching point where the sentinels are inserted. As the server executes the original program and reaches these data entrances, the current intermediate values are extracted and sent to the sentinel branches.

Once the client uploads the original inputs and the main program with the instrumented sentinels, the server starts the computation task. As the computation progresses and reaches the data entrance of the sentinels, the intermediate values $I_1$ and $I_2$ are extracted and sent to the respective sentinels. Then the check process begins. Both sentinels in FBC are designed as a $2n$-step process, comprising an n-step forward process and an n-step backward process. Each step in the forward process $\left\{f_1(\cdot), f_2(\cdot), \cdots, f_n(\cdot)\right\}$ should be a reversible arithmetic operation, composite operator, or function. The backward process is the reverse arrangement of the steps in the forward process, where each step is the inverse function of the corresponding step in the forward process. Mathematically, it can be expressed as: 
$$S_i^{\textit {Backward }}=f_i^{-1}(\cdot)$$

After the completion of the forward and backward process, the sentinels' computing is finished, and the outputs $R_1$ and $R_2$ are returned. In any accurate computing paradigm, it is expected that $I_1=R_1$ and $I_2=R_2$ because regardless of the specific operators used in each step, the forward process and the backward process are mutually reversible. In other words:
$$
\begin{aligned}
S(I) & =S^{\textit {Forward }} \cdot S^{\textit {Backward }}(I) \\
& =f_1 \cdot f_2 \cdots f_n \cdot f_n^{-1} \cdot f_{n-1}^{-1} \cdots \cdots f_1^{-1}(I)=I
\end{aligned}
$$
It is crucial to note that when detecting a floating-point program, we should consider the inherent truncation error of floating-point operations and relax the judgement equation as follows to avoid false positives:
$$
I - S^{\textit {Forward }} \cdot S^{\textit {Backward }}(I) < \delta
$$
where $\delta$ is a predefined small threshold used to distinguish between the normal system errors in floating-point computations and the approximation errors introduced by AC.

On the contrast, in the case of a malicious server who launched DHAC, AC errors are introduced and continuously accumulate during the computation process of the sentinels. This accumulation of errors can potentially result in a numerical discrepancy between $I_1$ and $R_1$, or $I_2$ and $R_2$. If any sentinel detects significant inconsistencies between $I$ and $R$, it will return a DHAC positive signal to alert the client. If all sentinels do not report any positive signal, the current result can be considered as an accurate computation result and can be directly used. 

\subsection{RCC vs FBC}

While FBC successfully addresses the limitation of RCC in terms of its applicability to integer programs only, both RCC and FBC possess their own advantages. In addition to their applicability, we have summarized the characteristics of these two DHAC detecting methods based on three other aspects. These characteristics are presented in Table \ref{table1}. 

\begin{table}[ht]
    \centering
    \caption{Comparison between RCC and FBC}
    \renewcommand\arraystretch{2}
    \label{table1}
    \begin{tabular}{|c|l|l|}
    \hline
    \textbf{Comparison} & \multicolumn{1}{c|}{\textbf{RCC}} & \multicolumn{1}{c|}{\textbf{FBC}} \\ \hline
    \textbf{Applicability} & Integer only & \textbf{Integer and floating point} \\ \hline
    \textbf{Obscurity} & \textbf{Imperceptible} & Perceptible \\ \hline
    \textbf{Independence} & \textbf{\begin{tabular}[c]{@{}l@{}}Independent check\\ rounds\end{tabular}} & \begin{tabular}[c]{@{}l@{}}Instrumented in the original\\ program\end{tabular} \\ \hline
    \textbf{Cost} & Local cost & \textbf{Small time cost} \\ \hline
    \end{tabular}
\end{table}

As we assumed, a malicious server has the ability to switch back from $P_{appx}$ to $P_{acc}$ when it suspects that the clients may be monitoring its behavior or checking the program. Thus, an overt checking behavior may trigger the malicious server’s alert to pause DHAC. In this regard, the check process of RCC is performed locally and remains imperceptible to the server. The server has no knowledge of how, when, or on which parts the check process is initiated. This provides an advantage to RCC as the server cannot anticipate or interfere with the RCC processes. On the other hand, FBC requires the client to instrument sentinel branches into the original program before uploading it to the server. These sentinel branches are specifically designed for DHAC checking purposes and are unrelated to the actual computing tasks. The presence of these sentinels may potentially expose the clients’ detection intent if the CSP thoroughly analyze the client's programs. However, in real world, the server may not check all clients’ programs, especially considering the large number of normal users among whom a client may be just one among thousands or millions. 

Furthermore, the check process of RCC is independent to the server’s side. Clients have the flexibility to pause, selectively skip, or adjust the number of check rounds as they want without any impact on the server's side. This gives clients full control over the checking process in RCC. In contrast, FBC needs to instrument sentinel branches into the program before the computing task begins. If clients need to change the location or adjust the number of sentinels, they must pause the server’s work, make the necessary adjustments to the program, and then reupload the instrumented version. This introduces interruption to the computing process.

An advantage of FBC is its cost-effectiveness. In FBC, the computing cost of the sentinel branches is borne by the server because the additional instrumented sentinel branches are computed on the server's side. For users, the potential cost is the possibility of increased computation time, as the server requires additional time to compute the sentinel branches. But compared to the main program, the computational workload of the sentinel branches is typically very small, and its impact on overall computation time is not significant. For on-demand cloud services, the increased computation time due to FBC may result in some additional rental costs. For cost models such as reserved-instances, which follow a monthly or annual billing cycle, the cost of FBC can be negligible. On the other hand, in RCC, the check process is completed locally by the client. Consequently, the client needs to afford the computing cost associated with the check processes, even though the check cost is significantly reduced compared to operations in the linear space. 

In summary, both RCC and FBC provide unique advantages in DHAC scenarios. RCC offers the advantage of being imperceptible to the server and provides more flexibility in terms of the timing and location of detection. FBC provides broader applicability by enabling the detection of DHAC in floating-point programs, and the cost of the check process is borne by the server rather than the clients. The choice between RCC and FBC depends on the specific requirements and constraints of the DHAC scenario. Both of them can detect DHAC attacks individually and are not mutually exclusive, allowing for their collaborative use.

\section{Experiments}

In this section, we focus on reporting the detection accuracy of RCC and FBC. We simulate a DHAC scenario with approximate hardware obtained from EvoApproxLib, as well as truncation-based floating-point operators. We then conducted tests to evaluate the detection accuracy of RCC and FBC in this simulated scenario. In subsection VI.A, we will provide a detailed description of the experiment setup, while in subsections VI.B and VI.C, we will present the detection accuracy results for RCC and FBC, respectively.

\subsection{Setup}

In our simulation, we considered two distinct AC environments: one for integer programs and another for floating-point programs. For integer programs, the binary data length is set to 16 bits. To simulate the approximate environment for integer programs, we utilized EvoApproxLib v1.1, which is a library that provides approximate hardware unit designs with various approximation degree and physical parameters \cite{evoapproxlib}. We deployed three approximate adders and three multipliers, each with different physical parameters. This allowed us to conduct tests with varying levels of approximation. 

For floating-point programs, we utilized a representation using 64-bit double precision numbers following the IEEE-754 format. The AC paradigm is built based on truncated adders and multipliers. Specifically, we utilized two truncation-based arithmetic units, where the mantissa part of the operands had 10 and 20 bits truncated, respectively. The remaining bits were still computed accurately.

The accurate paradigm, denoted as $P_{acc}$, executes the test programs using normal, accurate hardware. In contrast, the simulated DHAC scenario shifts the computation to the approximate paradigm $P_{appx}$, where the entire test programs are executed in the aforementioned approximate environment. To provide a comprehensive analysis, we evaluated and listed the power and area of the hardware units used in this experiment in Table \ref{table2}. The parameters of the EvoApproxLib units (the upper 8 units marked by * in the table) are directly referenced from their official website. The parameters of the floating-point arithmetic units are programmed in Verilog and synthesized using Cadence RTL Compiler with FreePDK 45nm library. All the test programs, as well as the two proposed detection methods (RCC and FBC), are implemented using Python v3.9.

\begin{table}[ht]
    \centering
    \caption{Hardware Parameters of Arithmetic Units}
    \renewcommand\arraystretch{1.4}
    \label{table2}
\begin{tabular}{ccccc}
\hline
\multicolumn{1}{|c|}{\textbf{Category}} & \multicolumn{1}{c|}{\textbf{Units}} & \multicolumn{1}{c|}{\textbf{\begin{tabular}[c]{@{}c@{}}Power\\  (mW)\end{tabular}}} & \multicolumn{1}{c|}{\textbf{\begin{tabular}[c]{@{}c@{}}Delay \\ (ns)\end{tabular}}} & \multicolumn{1}{c|}{\textbf{\begin{tabular}[c]{@{}c@{}}Area \\ (um2)\end{tabular}}} \\ \hline
\multicolumn{1}{|c|}{\multirow{4}{*}{\begin{tabular}[c]{@{}c@{}}16-bit \\ integer adders\end{tabular}}} & \multicolumn{1}{c|}{Accurate *} & \multicolumn{1}{c|}{0.072} & \multicolumn{1}{c|}{1.28} & \multicolumn{1}{c|}{141.7} \\ \cline{2-5} 
\multicolumn{1}{|c|}{} & \multicolumn{1}{c|}{add16u\_1DM *} & \multicolumn{1}{c|}{0.065} & \multicolumn{1}{c|}{1.12} & \multicolumn{1}{c|}{138.4} \\ \cline{2-5} 
\multicolumn{1}{|c|}{} & \multicolumn{1}{c|}{add16u\_0RN *} & \multicolumn{1}{c|}{0.06} & \multicolumn{1}{c|}{1.08} & \multicolumn{1}{c|}{115.9} \\ \cline{2-5} 
\multicolumn{1}{|c|}{} & \multicolumn{1}{c|}{add16u\_0Q7 *} & \multicolumn{1}{c|}{0.051} & \multicolumn{1}{c|}{0.95} & \multicolumn{1}{c|}{100.4} \\ \hline
\multicolumn{1}{|c|}{\multirow{4}{*}{\begin{tabular}[c]{@{}c@{}}16-bit \\ integer\\  multipliers\end{tabular}}} & \multicolumn{1}{c|}{Accurate *} & \multicolumn{1}{c|}{2.202} & \multicolumn{1}{c|}{3.11} & \multicolumn{1}{c|}{3203.0} \\ \cline{2-5} 
\multicolumn{1}{|c|}{} & \multicolumn{1}{c|}{mul16u\_60L *} & \multicolumn{1}{c|}{2.173} & \multicolumn{1}{c|}{3.13} & \multicolumn{1}{c|}{3052.3} \\ \cline{2-5} 
\multicolumn{1}{|c|}{} & \multicolumn{1}{c|}{mul16u\_0ZG *} & \multicolumn{1}{c|}{1.984} & \multicolumn{1}{c|}{3.23} & \multicolumn{1}{c|}{3094.1} \\ \cline{2-5} 
\multicolumn{1}{|c|}{} & \multicolumn{1}{c|}{mul16u\_GZ7 *} & \multicolumn{1}{c|}{1.859} & \multicolumn{1}{c|}{2.84} & \multicolumn{1}{c|}{2332.4} \\ \hline
\multicolumn{1}{|c|}{\multirow{3}{*}{\begin{tabular}[c]{@{}c@{}}64-bit \\ floating point \\ adders\end{tabular}}} & \multicolumn{1}{c|}{Accurate} & \multicolumn{1}{c|}{1.481} & \multicolumn{1}{c|}{8.15} & \multicolumn{1}{c|}{25825.1} \\ \cline{2-5} 
\multicolumn{1}{|c|}{} & \multicolumn{1}{c|}{10-bit truncation} & \multicolumn{1}{c|}{0.837} & \multicolumn{1}{c|}{6.73} & \multicolumn{1}{c|}{15334.8} \\ \cline{2-5} 
\multicolumn{1}{|c|}{} & \multicolumn{1}{c|}{20-bit truncation} & \multicolumn{1}{c|}{0.668} & \multicolumn{1}{c|}{5.35} & \multicolumn{1}{c|}{12442.8} \\ \hline
\multicolumn{1}{|c|}{\multirow{3}{*}{\begin{tabular}[c]{@{}c@{}}64-bit \\ floating point \\ multipliers\end{tabular}}} & \multicolumn{1}{c|}{Accurate} & \multicolumn{1}{c|}{10.831} & \multicolumn{1}{c|}{9.75} & \multicolumn{1}{c|}{29402.6} \\ \cline{2-5} 
\multicolumn{1}{|c|}{} & \multicolumn{1}{c|}{10-bit truncation} & \multicolumn{1}{c|}{6.674} & \multicolumn{1}{c|}{8.49} & \multicolumn{1}{c|}{20260.2} \\ \cline{2-5} 
\multicolumn{1}{|c|}{} & \multicolumn{1}{c|}{20-bit truncation} & \multicolumn{1}{c|}{3.595} & \multicolumn{1}{c|}{6.43} & \multicolumn{1}{c|}{12555.7} \\ \hline
\multicolumn{5}{l}{*: The parameters are directly referenced from \cite{evoapproxlib}}
\end{tabular}
\end{table}

\subsection{Evaluation on RCC}

To evaluate the detection accuracy of RCC, we deployed four programs: Euler method, Runge-Kutta method, Finite Input Response (FIR) Filter, and a 2×2 Convolution Kernel. The Euler method and Runge-Kutta method are classic numerical computing methods and are widely used in various scientfic and engineering applications. Each method involves a 10-step iteration with a random step size on polynomials of different orders. The FIR Filter is a well-known filter widely used in sensors, radar and machine vision. The 2×2 Convolution Kernel is a fundamental operator in image processing and computer vision. It performs a convolution operation between two 2×2 matrices. The numbers of basic operation components in each program are listed in Table \ref{table3}. For each program, we randomly generate 10000 integer input vectors as the input sets. As mentioned in Table \ref{table2}, the three AC adders and three AC multipliers can be combined into nine arithmetic unit combinations, each providing different degrees of approximation. In this experiment, RCC employs three check rounds, and the prime numbers used in the three rounds are settled as 3, 5 and 7. 

\begin{table}[ht]
\centering
\caption{Numbers of Basic Arithmetic Operations}
\renewcommand\arraystretch{1.4}
\label{table3}
\begin{tabular}{|c|c|ccc|}
\hline
\multirow{2}{*}{\textbf{Program}} & \multirow{2}{*}{\textbf{Order}} & \multicolumn{3}{c|}{\textbf{Number of Nodes}} \\ \cline{3-5} 
 &  & \multicolumn{1}{c|}{\textbf{Add \& Sub}} & \multicolumn{1}{c|}{\textbf{Mul}} & \textbf{Total} \\ \hline
\multirow{2}{*}{Euler Method} & 2 & \multicolumn{1}{c|}{32} & \multicolumn{1}{c|}{22} & 54 \\ \cline{2-5} 
 & 3 & \multicolumn{1}{c|}{43} & \multicolumn{1}{c|}{45} & 88 \\ \hline
\multirow{2}{*}{Runge-Kutta Method} & 2 & \multicolumn{1}{c|}{42} & \multicolumn{1}{c|}{32} & 74 \\ \cline{2-5} 
 & 3 & \multicolumn{1}{c|}{73} & \multicolumn{1}{c|}{75} & 148 \\ \hline
FIR Filter & \textbackslash{} & \multicolumn{1}{c|}{10} & \multicolumn{1}{c|}{11} & 21 \\ \hline
2X2 Convolution Kernel & \textbackslash{} & \multicolumn{1}{c|}{3} & \multicolumn{1}{c|}{4} & 7 \\ \hline
\end{tabular}
\end{table}

\begin{figure*}
    \centering
    \includegraphics [scale=0.28] {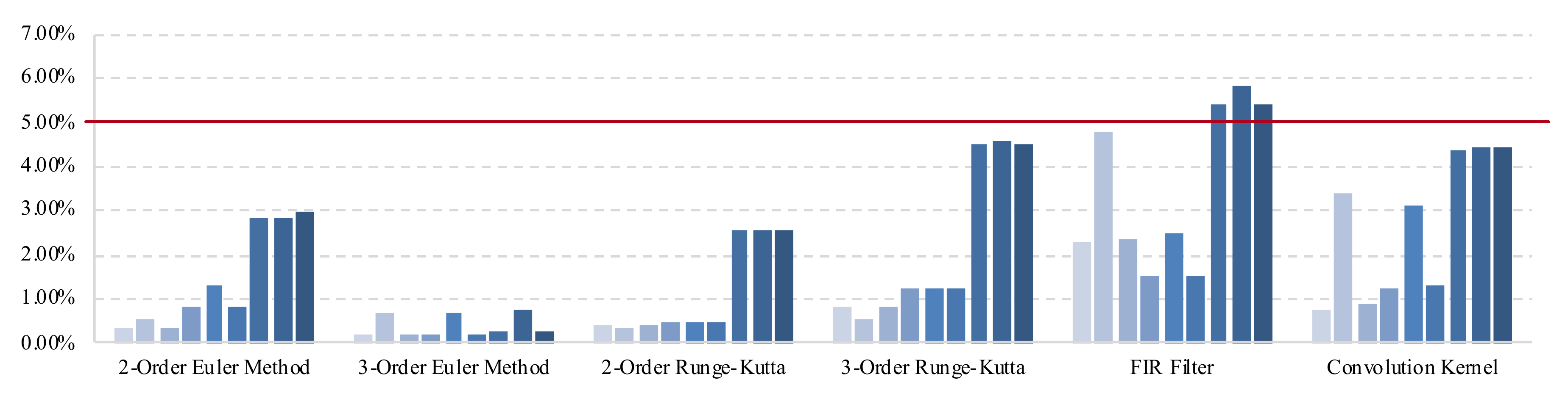}
    \caption{The Mean Relative Errors (\%) of The Test Programs with Different AC Configurations (From the left side in each group, the nine adder and multiplier combinations are: (1DM, 60L), (1DM, 0ZG), (1DM, GZ7), (0RN, 60L), (0RN, 0ZG), (0RN, GZ7), (0Q7, 60L), (0Q7, 0ZG), (0Q7, GZ7))}
    \label{fig:The Mean Relative Errors}
\end{figure*}

\begin{figure}
    \includegraphics [scale=0.62] {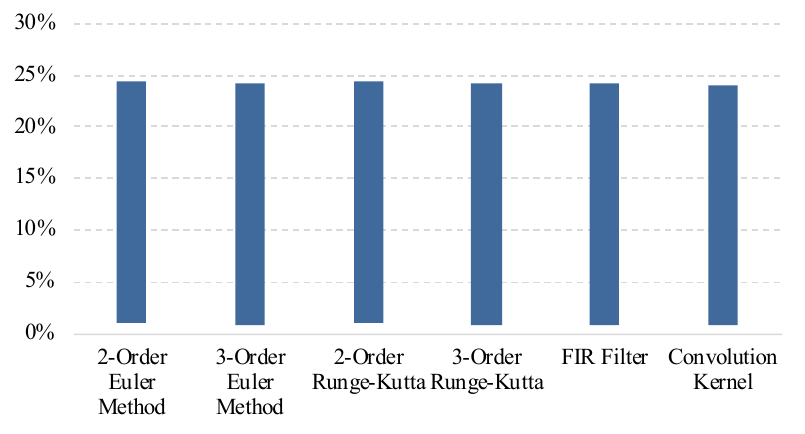}
    \caption{The Energy Savings of the Test Programs}
    \label{fig:The Energy Savings of the Test Programs}
\end{figure}

To ensure the validity of the simulated AC environment, we first conduct an environment to evaluate the output degradation to confirm that the approximation errors are within a reasonable range, and did not lead to intuitively incorrect results. The mean relative error (MRE) of the test programs using the nine different AC unit combinations is presented in Fig. \ref{fig:The Mean Relative Errors}. Overall, with the exception of the FIR Filter, which exhibits errors exceeding 5\% in the three most aggressive approximation configurations, the errors of all four programs remained within 5\% across all AC configurations. This indicates that the simulated approximation environment generally ensures reasonable results, with deviations that did not significantly affect the overall correctness of the computations. The energy saving of the test programs with different AC configurations are shown in Fig. \ref{fig:The Energy Savings of the Test Programs}, where each bar in this chart displays the maximum and minimum values of energy saving for nine different AC configurations. The highest energy saving achieved by the AC adder and multiplier combinations is about 24\%. Here we did not employ more aggressive approximation strategies because the main goal of our experiment is to evaluate the DHAC detection accuracy of RCC, particularly in the cases with lower levels of approximation.

Table \ref{table4} presents the average detection accuracy of RCC across various AC configurations. It is worth noting that certain DHAC examples with specific inputs are undetectable because their outputs have no error when compared to accurate results, even though they are executed in an approximate computing environment. The occurrence of these no-error approximations is reasonable because not all combinations of operands in typical approximate integer hardware will result in errors. Another reason for some undetectable examples is the cancellation of errors during the computation process. In certain cases, AC introduced both positive and negative errors, and they happened to offset each other, resulting in no observable error in the final result. These undetectable DHAC cannot be identified solely at the data level. For all four test programs, RCC achieves a detection accuracy of over 98\% after the second check round and can detect over 99.8\% of detectable DHAC attack scenarios after completing all the three check rounds. The DHAC negative signal is more trustworthy when the RCC process involves more check rounds. The main reason for the detectable but not detected examples (False Negatives) is that the DHAC output still falls within the same residual class as the accurate results, despite containing approximation errors. We conducted comparative experiments in the accurate paradigm using the same inputs, and RCC did not produce any false-positive examples.

\begin{table*}[ht]
\centering
\caption{The Detection Accuracy (\%) of RCC on Integer Programs}
\renewcommand\arraystretch{1.2}
\label{table4}
\begin{tabular}{cccccccccc}
\hline
\multicolumn{1}{|c|}{\multirow{3}{*}{\textbf{Program}}} &
  \multicolumn{1}{c|}{\multirow{3}{*}{\textbf{Order}}} &
  \multicolumn{1}{c|}{\multirow{3}{*}{\textbf{Detectable*}}} &
  \multicolumn{7}{c|}{\textbf{RCC}} \\ \cline{4-10} 
\multicolumn{1}{|c|}{} &
  \multicolumn{1}{c|}{} &
  \multicolumn{1}{c|}{} &
  \multicolumn{1}{c|}{\textbf{Round 1}} &
  \multicolumn{1}{c|}{\begin{tabular}[c]{@{}c@{}}\textbf{Round 1}\\ \textbf{/Detectable}\end{tabular}} &
  \multicolumn{1}{c|}{\textbf{Round 2}} &
  \multicolumn{1}{c|}{\begin{tabular}[c]{@{}c@{}}\textbf{Round 2}\\ \textbf{/Detectable}\end{tabular}} &
  \multicolumn{1}{c|}{\textbf{Round 3}} &
  \multicolumn{1}{c|}{\begin{tabular}[c]{@{}c@{}}\textbf{Round 3}\\ \textbf{/Detectable}\end{tabular}} &
  \multicolumn{1}{c|}{\textbf{False Positive}} \\ \hline
\multicolumn{1}{|c|}{\multirow{2}{*}{Euler Method}} &
  \multicolumn{1}{c|}{2} &
  \multicolumn{1}{c|}{93.84} &
  \multicolumn{1}{c|}{76.89} &
  \multicolumn{1}{c|}{81.93} &
  \multicolumn{1}{c|}{93.28} &
  \multicolumn{1}{c|}{99.4} &
  \multicolumn{1}{c|}{93.83} &
  \multicolumn{1}{c|}{\textbf{99.98}} &
  \multicolumn{1}{c|}{0} \\ \cline{2-10} 
\multicolumn{1}{|c|}{} &
  \multicolumn{1}{c|}{3} &
  \multicolumn{1}{c|}{77.64} &
  \multicolumn{1}{c|}{63.20} &
  \multicolumn{1}{c|}{81.4} &
  \multicolumn{1}{c|}{76.58} &
  \multicolumn{1}{c|}{98.6} &
  \multicolumn{1}{c|}{77.61} &
  \multicolumn{1}{c|}{\textbf{99.96}} &
  \multicolumn{1}{c|}{0} \\ \hline
\multicolumn{1}{|c|}{\multirow{2}{*}{\begin{tabular}[c]{@{}c@{}}Runge-Kutta\\ Method\end{tabular}}} &
  \multicolumn{1}{c|}{2} &
  \multicolumn{1}{c|}{84.77} &
  \multicolumn{1}{c|}{65.79} &
  \multicolumn{1}{c|}{77.61} &
  \multicolumn{1}{c|}{83.61} &
  \multicolumn{1}{c|}{98.63} &
  \multicolumn{1}{c|}{84.76} &
  \multicolumn{1}{c|}{\textbf{99.98}} &
  \multicolumn{1}{c|}{0} \\ \cline{2-10} 
\multicolumn{1}{|c|}{} &
  \multicolumn{1}{c|}{3} &
  \multicolumn{1}{c|}{88.63} &
  \multicolumn{1}{c|}{72.64} &
  \multicolumn{1}{c|}{81.95} &
  \multicolumn{1}{c|}{86.88} &
  \multicolumn{1}{c|}{98.03} &
  \multicolumn{1}{c|}{88.53} &
  \multicolumn{1}{c|}{\textbf{99.88}} &
  \multicolumn{1}{c|}{0} \\ \hline
\multicolumn{1}{|c|}{FIR Filter} &
  \multicolumn{1}{c|}{\textbackslash{}} &
  \multicolumn{1}{c|}{98.82} &
  \multicolumn{1}{c|}{79.47} &
  \multicolumn{1}{c|}{80.4} &
  \multicolumn{1}{c|}{97.98} &
  \multicolumn{1}{c|}{99.15} &
  \multicolumn{1}{c|}{98.81} &
  \multicolumn{1}{c|}{\textbf{99.98}} &
  \multicolumn{1}{c|}{0} \\ \hline
\multicolumn{1}{|c|}{\begin{tabular}[c]{@{}c@{}}Convolution\\ Kernel\end{tabular}} &
  \multicolumn{1}{c|}{\textbackslash{}} &
  \multicolumn{1}{c|}{89.67} &
  \multicolumn{1}{c|}{77.33} &
  \multicolumn{1}{c|}{86.2} &
  \multicolumn{1}{c|}{89.39} &
  \multicolumn{1}{c|}{99.68} &
  \multicolumn{1}{c|}{89.66} &
  \multicolumn{1}{c|}{\textbf{99.98}} &
  \multicolumn{1}{c|}{0} \\ \hline
\multicolumn{10}{l}{\begin{tabular}[c]{@{}l@{}}* Some examples are undetectable in principle because with certain inputs, the DHAC output is equals to the accurate result.   The approximate \\ computing components introduce no computation errors.\end{tabular}}
\end{tabular}
\end{table*}

\subsection{Detection on Floating-point Programs}

\begin{figure}
    \includegraphics[scale=0.75]{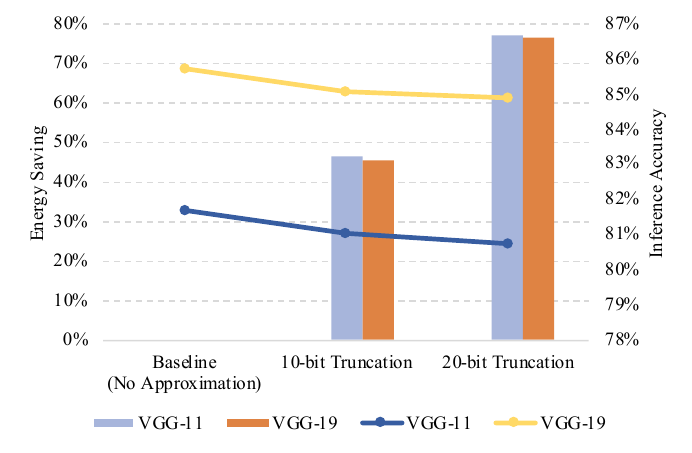}
    \caption{Inference Accuracy and Energy Saving of Different AC Configurations}
    \label{fig:Inference Accuracy and Energy Saving of Different AC Configurations}
\end{figure}

We select VGG-11 and VGG-19, two representative convolutional neural network models in the deep learning field, as the test program to evaluate the DHAC detection capability of FBC. Both of the two models are trained using the Cifar-10 data set, and we simulated scenarios where DHAC attacks occurred during both the training and inference processes. We build an approximate computing environment based on truncation, a popular approximation method in floating-point programs. Specifically, we set the truncated mantissa bits to 10 and 20 bits. The energy saving and inference accuracy are presented in Fig. \ref{fig:Inference Accuracy and Energy Saving of Different AC Configurations}. The bars with the left y-axis in this figure represent the energy saving of VGG-11 and VGG-19. A taller bar means a better optimization on energy consumption. The baseline is the case without any approximation and hence no energy saving. The line with the right y-axis in the graph represents the decrease in inference accuracy as the level of approximation increases. For both VGG-11 and VGG-19, we observed only a small degradation of less than 1\% in inference accuracy, while achieving energy savings of 46\% and 77\% for 10- and 20-bit truncation, respectively.

We design three types of sentinel branches. The first type is the addition sentinel, which consists of three consecutive additions in the forward process and three subtractions in the backward process. The second type is the multiplication sentinel, which follows the same structure but employs multiplications and divisions instead. These two sentinels can also identify whether the attacker solely deploys approximate adders or multipliers. The third type is a tan-arctan branch, which can be seen as a mixture of additions and multiplications.

\begin{table*}[ht]
\centering
\caption{The Detection Accuracy (\%) of FBC on Floating Point Programs with 10-bit Truncation}
\renewcommand\arraystretch{1.2}
\label{table5}
\begin{tabular}{|ccc|cccc|cccc|}
\hline
\multicolumn{1}{|c|}{\multirow{2}{*}{\textbf{Training or Inference}}} & \multicolumn{1}{c|}{\multirow{2}{*}{\textbf{Program}}} & \multirow{2}{*}{\textbf{Epoch}} & \multicolumn{4}{c|}{\textbf{64 channel conv. layer}} & \multicolumn{4}{c|}{\textbf{128 channel conv. layer}} \\ \cline{4-11} 
\multicolumn{1}{|c|}{} & \multicolumn{1}{c|}{} &  & \multicolumn{1}{c|}{add.} & \multicolumn{1}{c|}{mul.} & \multicolumn{1}{c|}{tan} & False Positive & \multicolumn{1}{c|}{add.} & \multicolumn{1}{c|}{mul.} & \multicolumn{1}{c|}{tan} & False Positive \\ \hline
\multicolumn{1}{|c|}{\multirow{4}{*}{Training}} & \multicolumn{1}{c|}{\multirow{2}{*}{VGG-11}} & 25 & \multicolumn{1}{c|}{98.85} & \multicolumn{1}{c|}{98.65} & \multicolumn{1}{c|}{99.76} & 0 & \multicolumn{1}{c|}{98.88} & \multicolumn{1}{c|}{97.68} & \multicolumn{1}{c|}{99.84} & 0 \\ \cline{3-11} 
\multicolumn{1}{|c|}{} & \multicolumn{1}{c|}{} & 50 & \multicolumn{1}{c|}{98.85} & \multicolumn{1}{c|}{98.88} & \multicolumn{1}{c|}{99.88} & 0 & \multicolumn{1}{c|}{98.92} & \multicolumn{1}{c|}{98.06} & \multicolumn{1}{c|}{99.8} & 0 \\ \cline{2-11} 
\multicolumn{1}{|c|}{} & \multicolumn{1}{c|}{\multirow{2}{*}{VGG-19}} & 25 & \multicolumn{1}{c|}{98.79} & \multicolumn{1}{c|}{97.79} & \multicolumn{1}{c|}{99.7} & 0 & \multicolumn{1}{c|}{99.02} & \multicolumn{1}{c|}{95.82} & \multicolumn{1}{c|}{99.7} & 0 \\ \cline{3-11} 
\multicolumn{1}{|c|}{} & \multicolumn{1}{c|}{} & 50 & \multicolumn{1}{c|}{98.75} & \multicolumn{1}{c|}{97.85} & \multicolumn{1}{c|}{99.7} & 0 & \multicolumn{1}{c|}{98.72} & \multicolumn{1}{c|}{95.56} & \multicolumn{1}{c|}{99.58} & 0 \\ \hline
\multicolumn{1}{|c|}{\multirow{2}{*}{Inference}} & \multicolumn{2}{c|}{VGG-11} & \multicolumn{1}{c|}{99.13} & \multicolumn{1}{c|}{91.4} & \multicolumn{1}{c|}{99.98} & 0 & \multicolumn{1}{c|}{99.3} & \multicolumn{1}{c|}{92.87} & \multicolumn{1}{c|}{100} & 0 \\ \cline{2-11} 
\multicolumn{1}{|c|}{} & \multicolumn{2}{c|}{VGG-19} & \multicolumn{1}{c|}{99.22} & \multicolumn{1}{c|}{96.42} & \multicolumn{1}{c|}{99.94} & 0 & \multicolumn{1}{c|}{98.88} & \multicolumn{1}{c|}{98.9} & \multicolumn{1}{c|}{99.33} & 0 \\ \hline
\multicolumn{3}{|c|}{\textbf{Average}} & \multicolumn{1}{c|}{\textbf{98.93}} & \multicolumn{1}{c|}{\textbf{96.83}} & \multicolumn{1}{c|}{\textbf{99.83}} & \textbf{0} & \multicolumn{1}{c|}{\textbf{98.95}} & \multicolumn{1}{c|}{\textbf{96.48}} & \multicolumn{1}{c|}{\textbf{99.71}} & \textbf{0} \\ \hline
\end{tabular}
\end{table*}

\begin{table*}[ht]
\centering
\caption{The Detection Accuracy (\%) of FBC on Floating Point Programs with 20-bit Truncation}
\label{table6}
\renewcommand\arraystretch{1.2}
\begin{tabular}{|ccc|cccc|cccc|}
\hline
\multicolumn{1}{|c|}{\multirow{2}{*}{\textbf{Training or Inference}}} & \multicolumn{1}{c|}{\multirow{2}{*}{\textbf{Program}}} & \multirow{2}{*}{\textbf{Epoch}} & \multicolumn{4}{c|}{\textbf{64 channel conv. layer}} & \multicolumn{4}{c|}{\textbf{128 channel conv. layer}} \\ \cline{4-11} 
\multicolumn{1}{|c|}{} & \multicolumn{1}{c|}{} &  & \multicolumn{1}{c|}{add.} & \multicolumn{1}{c|}{mul.} & \multicolumn{1}{c|}{tan} & False Positive & \multicolumn{1}{c|}{add.} & \multicolumn{1}{c|}{mul.} & \multicolumn{1}{c|}{tan} & False Positive \\ \hline
\multicolumn{1}{|c|}{\multirow{4}{*}{Training}} & \multicolumn{1}{c|}{\multirow{2}{*}{VGG-11}} & 25 & \multicolumn{1}{c|}{99.95} & \multicolumn{1}{c|}{99.87} & \multicolumn{1}{c|}{99.86} & 0 & \multicolumn{1}{c|}{99.97} & \multicolumn{1}{c|}{99.79} & \multicolumn{1}{c|}{99.92} & 0 \\ \cline{3-11} 
\multicolumn{1}{|c|}{} & \multicolumn{1}{c|}{} & 50 & \multicolumn{1}{c|}{99.99} & \multicolumn{1}{c|}{99.91} & \multicolumn{1}{c|}{99.94} & 0 & \multicolumn{1}{c|}{99.96} & \multicolumn{1}{c|}{99.78} & \multicolumn{1}{c|}{99.92} & 0 \\ \cline{2-11} 
\multicolumn{1}{|c|}{} & \multicolumn{1}{c|}{\multirow{2}{*}{VGG-19}} & 25 & \multicolumn{1}{c|}{99.97} & \multicolumn{1}{c|}{99.81} & \multicolumn{1}{c|}{99.9} & 0 & \multicolumn{1}{c|}{99.98} & \multicolumn{1}{c|}{99.7} & \multicolumn{1}{c|}{99.74} & 0 \\ \cline{3-11} 
\multicolumn{1}{|c|}{} & \multicolumn{1}{c|}{} & 50 & \multicolumn{1}{c|}{99.99} & \multicolumn{1}{c|}{99.78} & \multicolumn{1}{c|}{99.9} & 0 & \multicolumn{1}{c|}{99.95} & \multicolumn{1}{c|}{99.55} & \multicolumn{1}{c|}{99.74} & 0 \\ \hline
\multicolumn{1}{|c|}{\multirow{2}{*}{Inference}} & \multicolumn{2}{c|}{VGG-11} & \multicolumn{1}{c|}{99.98} & \multicolumn{1}{c|}{99.98} & \multicolumn{1}{c|}{100} & 0 & \multicolumn{1}{c|}{99.97} & \multicolumn{1}{c|}{99.99} & \multicolumn{1}{c|}{100} & 0 \\ \cline{2-11} 
\multicolumn{1}{|c|}{} & \multicolumn{2}{c|}{VGG-19} & \multicolumn{1}{c|}{99.99} & \multicolumn{1}{c|}{99.98} & \multicolumn{1}{c|}{100} & 0 & \multicolumn{1}{c|}{99.96} & \multicolumn{1}{c|}{99.87} & \multicolumn{1}{c|}{100} & 0 \\ \hline
\multicolumn{3}{|c|}{\textbf{Average}} & \multicolumn{1}{c|}{\textbf{99.97}} & \multicolumn{1}{c|}{\textbf{99.89}} & \multicolumn{1}{c|}{\textbf{99.93}} & \textbf{0} & \multicolumn{1}{c|}{\textbf{99.96}} & \multicolumn{1}{c|}{\textbf{99.78}} & \multicolumn{1}{c|}{\textbf{99.89}} & \textbf{0} \\ \hline
\end{tabular}
\end{table*}

We instrument these two sentinels into the 64-channel convolutional layer and the 128-channel convolutional layer of the VGG-11 and VGG-19 models. When the computation reaches the data entrance of the sentinels, the intermediate value is extracted and used as the input for the corresponding sentinel branch to initiate the checking process. For addition and multiplication sentinels, the other three operands involved in the continuous three additions or multiplications are randomly generated within the range of 0 to 1. It is worth noting that even if all the required operands for the sentinel branch can be derived from intermediate values of the program, we still recommend including randomly generated floating-point numbers with long mantissa bits. This step can eliminate any data-related features of the intermediate values before entering the sentinel branches. For the tan-arctan branch, no additional input value is generated because it only requires one input.

The detection experiment is repeated for 10000 times, with each time using different intermediate values as the starting point of the sentinels. The detection accuracy for the 10-bit and 20-bit truncation scenarios is presented in Table \ref{table5} and Table \ref{table6}, respectively. For the 10-bit truncated programs, the detection accuracy of the addition sentinels reaches 98.9\%, while the average detection accuracy for multiplication and tan-arctan sentinels exceeds 96.4\% and 99.7\%. The detection accuracy of these three sentinels all improves to over 99.7\% when it turns to 20-bit truncation scenario, which indicates that a more aggressive approximation strategy will be easier to detect. Two main reasons lead to the undetected DHAC examples (False Negatives). In rare cases, most of the mantissa bits are 0, resulting in no truncation error being exposed during the computation in the sentinel branch. Another scenario occurs when the positive errors and negative errors in the sentinel branch’s computation offset each other, resulting in the final result being very close to the starting value. The comparative experiments in the accurate paradigm show that all the sentinels did not produce any false-positive examples.

\begin{figure}
    \centering
    \includegraphics[scale=0.75] {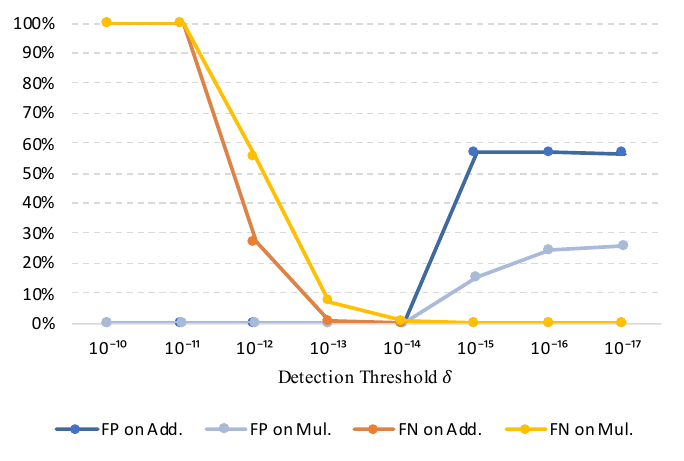}
    \caption{False Positive (FP) and False Negative (FN) Rate}
    \label{fig:Figure_floating_threshold}
\end{figure}

We further analyzed how the detection threshold $\delta$ influence the false positive (FP) and false negative (FN) rate of FBC. Intuitively, in floating-point programs, there can be small truncation errors due to the inherent principles of floating-point computation, and AC will introduce more errors. Therefore, selecting an appropriate detection threshold is crucial for clients to distinguish between AC errors and normal floating-point truncation errors. Fig. \ref{fig:Figure_floating_threshold} indicates the variations in FP and FN rates as the threshold value decreases. An overlarge $\delta$ cannot only point out AC errors and an very small $\delta$ will be too strict that normal floating-point truncation errors are not tolerated. A threshold $\delta$ between $[10^{-14}, 10^{-13}]$ can achieve a high DHAC detection rate while holding a low FP rate. This experiment also demonstrates that AC error exhibit an order of magnitude difference compared to normal floating-point truncation errors. This difference can be effectively recognized by setting appropriate threshold values.

\section{Conclusions}

In this paper, we have delved into the dark side behind the widespread adoption of AC and proposed a malicious misusing scenario called DHAC. This malicious attack can be easily launched by dishonest CSPs to gain illegal financial benefits. As a result, clients suffer from financial losses and experience degraded computing results because of DHAC. To address this issue, we have proposed two golden model free detection methods, namely RCC and FBC, to provide clients a low-cost weapon to continuously monitor candidate servers over an extended period. Our experimental results demonstrate that both RCC and FBC can achieve a detection accuracy of over 96\%-99\% DHAC instances without any misjudgment of legitimate accurate results. In our future work, we plan to extend the investigation of DHAC to other practical scenarios, such as Software-as-a-Service (SaaS) clouds and local black-box devices. By exploring these contexts, we aim to provide further insights and develop effective countermeasures against DHAC attacks.





\bibliographystyle{IEEEtranS}
\bibliography{main}

\end{document}